\documentstyle[aps, epsfig]{revtex}

\draft

\begin{document}

\twocolumn[
\hsize\textwidth\columnwidth\hsize\csname@twocolumnfalse\endcsname

\title{Friedel Oscillations and Charge-Density Waves Pinning in 
Quasi-one-dimensional Conductors: An X-ray Access.}
\author{S. Rouzi\`{e}re$^{1}$, S. Ravy$^{2}$, J.-P. Pouget$^{2}$ and 
S. Brazovskii$^{3}$}
\address{$^{1}$School of Chemistry, Physics and Environmental
Science, University of Sussex, Falmer, Brighton BN1 9QJ, UK}
\address{$^{2}$Laboratoire de physique des solides, CNRS UMR 8502, 
B\^{a}t. 510, Universit\'{e} Paris-sud, 91405 Orsay Cedex, France}
\address{$^{3}$Laboratoire de physique th\'{e}orique et mod\`{e}les 
statistiques, 
CNRS UMR 8626,  B\^{a}t. 100,
Universit\'{e} Paris-sud, 91405 Orsay Cedex, France}
\maketitle

\begin{abstract}
We present an x-ray diffraction study of the vanadium-doped blue bronze 
K$_{0.3}($Mo$_{0.972}$V$_{0.028}$)O$_{3}$. At low temperature, we have observed 
both an intensity asymmetry of the $\pm 2k_{F}$ satellite reflections 
relative to the pure compound, and a profile asymmetry of each satellite 
reflections. We show that the profile asymmetry is due to Friedel oscillation 
around the V substituant and that the intensity asymmetry is related to the 
charge density wave (CDW) pinning. These two effects, intensity and profile
asymmetries, give for the first time access to the local properties of CDW 
in disordered systems, including the pinning and even the phase shift of 
Friedel oscillations. 
\end{abstract}

\twocolumn]

\narrowtext

The effects of impurities on the physical properties of crystals is one of the
major issues of solid state physics. In metals, impurities are known to 
originate Friedel oscillations\cite {Friedel} (FOs) of the electronic density 
and to affect generic ones like charge-density waves (CDWs)\cite{CDW}. 
In quasi-one dimensional (1D) metals both features occur. 
Due to its divergent susceptibility at twice the Fermi
wave vector $2k_{F}$, a 1D electron gas is unstable with
respect to the formation of a $2k_{F}$-CDW, which, coupled to the lattice, 
induces a $2k_{F}$-periodic lattice distortion.
The $2k_{F}$-satellite reflections arising from this distortion provides a
rich variety of information on the CDW. 
While the peak intensity and its position give the amplitude and the wave 
length of the CDW, the satellite reflections profiles is related to the 
phase-phase correlation function of the CDW.
In CDW materials like blue bronze K$_{0.3}$MoO$_{3}$ \cite{Girault} or 
transition metal trichalcogenides like NbSe$_{3}$ \cite{Sweet},\cite{Rouz}, 
diffraction experiments have clearly demonstrated the loss of the CDW long 
range order upon doping. 
But despite this recent progress in the description of the
CDW structure, the microscopic nature of the pinning is still largely
unknown. 
Recently, two new methods have been found to access the local properties of 
CDW in disordered systems \cite{Ravy1} \cite{Ravy2}. The first one is the 
analysis of the intensity asymmetry (IA) of the satellite reflections at 
${\bf +}2{\bf k}_{F}/{\bf -}2{\bf k}_{F}$ from a Bragg reflection, which gives 
information of the value of the phase at an impurity position. 
Another method, that we describe here for
the first time, consists in analysing the profile asymmetry (PA) of 
the individual peaks. 
In this letter, we shall demonstrate that both effects, intensity and profile
asymmetries, can be observed simultaneously and that they provide a unique tool 
to access the FO's and the pinning. 

Let us first consider the general case of a modulated crystal containing 
impurities in concentration $c$. The substitution disorder is characterized 
by $\sigma ({\bf r})$, which is $1$ if an impurity sits in position 
${\bf r}$ and $0$ otherwise\cite {Krivo}. The displacements of the atoms from 
their regular positions are ${\bf u(r)}$. The total diffracted intensity at 
the scattering vector ${\bf Q=G}_{hkl}{\bf +q}$ close to one of the reciprocal
wave vector ${\bf G}_{hkl}$ reads, \cite{Ravy2}:

\begin{eqnarray}  \label{scatt}
&&I({\bf Q})= I_{L} + I_{d} + I_{a} \\
&&=\Delta f^{2}\left\langle \sigma _{{\bf q}}\sigma _{-{\bf q}%
}\right\rangle +\overline{f}^{2}\left\langle \left| {\bf Q.u}_{{\bf q}%
}\right| ^{2}\right\rangle%
-2\overline{f}\Delta f \mathop{\rm Im} \left\langle \sigma _{-{\bf q}}%
{\bf Q.u}_{{\bf q}}\right\rangle  \nonumber
\end{eqnarray}
where $\left\langle ..\right\rangle $ denotes an average on both disorder
and thermal fluctuations. $\Delta f=f_{I}-f$ is the difference between the
impurity and the host atom form factors, $\overline{f}=cf_{I}+(1-c)f$ is the
average form factor and ${\bf u}_{{\bf q}}$ and $\sigma _{{\bf q}}$ are the
Fourier transforms of ${\bf u(r)}$ and $\sigma ({\bf r})-c$ respectively.
${\bf q}$, which is any wave vector of the first Brillouin zone, is close to 
$2{\bf k}_{F}$ in the present case.

The first term gives the scattering intensity due to disorder,
usually called the ''Laue scattering''. The second term represents
the intensity due to atomic displacements {\it i.e.} the usual Fourier 
transform of the correlation function 
$\left\langle {\bf u(}0{\bf )u(r)}\right\rangle $.
The third term arises from the coupling between the disorder and
the displacements. Remarkably, $I_{a}$ is {\it linear} in ${\bf u}_{{\bf q}}$
and {\it odd} in ${\bf q}$, which changes the relative intensities at ${\bf G%
}_{hkl}{\bf +q}$ and ${\bf G}_{hkl}{\bf -q}$ with respect to that of the
pure crystal. An intensity asymmetry occurs around the ${\bf G}_{hkl}$
reciprocal position. Physically, this term is due to an interference between
the reference wave scattered by the impurity and the wave scattered by the
atomic displacements originated by the same impurity. In that sense, this
phenomenon can be seen as an {\it holography} of the lattice distortion around 
the impurity. 
$I_{a}$ cancels out when there is no coupling between the impurity 
positions and the atomic displacements ($\left\langle \sigma _{-{\bf q}}{\bf %
Q.u}_{{\bf q}}\right\rangle =\left\langle \sigma _{-{\bf q}}\right\rangle
\left\langle {\bf Q.u}_{{\bf q}}\right\rangle =0$). However such a coupling
does exist for FO's and CDW when it is pinned. In a simple model of strong
pinning, where the lattice distortion has an amplitude ${\bf u}_{2{\bf k}_{F}}$
and a phase $\varphi _{0}$ at the impurity site, it has been shown that 
\cite{Ravy1}, \cite{Ravy2}:

\begin{equation}
I_{a}({\bf Q}={\bf G}_{hkl}{\bf \pm }2{\bf k}_{F})\sim \mp \overline{f}%
\Delta f({\bf Q.u}_{2{\bf k}_{F}})\cos \varphi _{0}.  
\label{cosphi}
\end{equation}

This gives an intensity asymmetry of the ${\bf +}2{\bf k}_{F}/{\bf -}2%
{\bf k}_{F}$ satellite reflections and provides a unique way of evaluating 
$\cos \varphi _{0}$. In organic materials, ''white lines'' due to strong 
intensity asymmetry have been observed and used to determine the phase 
$\varphi _{0}$\cite{Ravy1},\cite{Brazo1}. 
With the same formalism, it can be shown that if the phase of 
the CDW undergoes a deformation around the defect, a profile asymmetry 
of each satellite is produced\cite{Ravy2}. 
The same effects is produced by the notorious phse shifts of FO's. 
In this communication we report on how these ideas have been tested in doped 
blue bronzes, which can be considered, with NbSe$_3$ as a prototype of 
charge-density wave materials.

\begin{figure}
\centerline{\epsfig{file=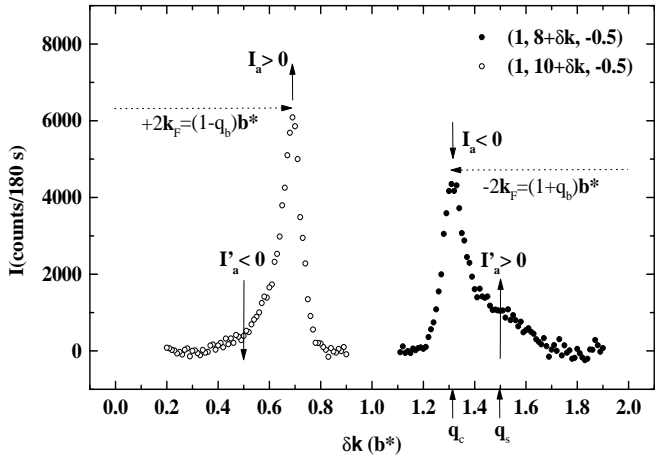,angle=90.0,height=6cm,width=8.5cm}}
\vskip 0.2cm
\caption{ High resolution profiles of the large-$k$ $(1,9+q_{b},6.5)$
and $(1,11-q_{b},6.5)$ peaks at 60 K. The position of the ${\bf q}_{c}$- and 
${\bf q}_{s}$-scattering are indicated on the x-axis. The vertical arrows
indicate the intensity asymmetry of the ${\bf q}_{c}$- and 
${\bf q}_{s}$-scattering. } 
\end{figure}

Quasi-one-dimensional blue bronze K$_{0.3}$MoO$_{3}$ undergoes a Peierls
transition at $T_{p}$=183K\cite{Schlenker}. Below $T_{p}$, satellite peaks
appear at the reduced wavevector ${\bf q}_{c}$, with ${\bf q}%
_{c}=(1,0.748,0.5)$ at 15K \cite{Poug1}. In order to maximize the effect of
impurities on the scattering, we have investigated heavily vanadium-doped blue
bronzes (2.8 at.\% V). X-ray diffraction studies were
performed with a three-circle diffractometer on a 12 kW rotating anode ($%
\lambda_{MoK_{a}}$=0.711 \AA ) and with a four-circle diffractometer at the
BM2 beamline in ESRF for high resolution measurements ($\lambda $=0.7\AA\
and $\lambda $=0.8\AA ). Previous studies had shown that at low temperature,
the $2k_{F}$-satellite reflections observed at the reduced position ${\bf q}%
_{c}$=($1,1-q_{b}=0.68,0.5$) were broadened, indicating the lost of the CDW
long range order \cite{Girault}. At last, the measured $%
2k_{F}=(1-q_{b})b^{*} $ ($b^{*}=2\pi /b$) was in accordance with the $2k_{F}$
value expected from the change in band filling, due to the substitution of
2.8\% V$^{5+}$ for Mo$^{6+}$\cite{Girault}.

 In order to gain evidence for an intensity asymmetry, we have measured the 
profiles of $2k_{F}/-2k_{F}$ scattering at different reciprocal 
positions and carefully compared the results obtained in doped and pure 
crystals. Typical high resolution scans in the chain (${\bf b}$) direction 
around two satellite reflections are shown in Fig. 1, after background 
substraction.
A casual inspection of Fig. 1 clearly reveals a profile asymmetry 
of the satellite reflections, already noted in Ref. \cite{Girault}. 
This asymmetric peaks can be 
analyzed by a sum of two components, a broad one, located at the reduced 
position ${\bf q}_{s}=(1,1-q_{s},0.5)$, where $q_{s}$=0.5 at 15 K, and a 
narrow one, corresponding to the regular ${\bf q}_{c}$-scattering by the CDW. 
 
\begin{figure}
\centerline{\epsfig{file=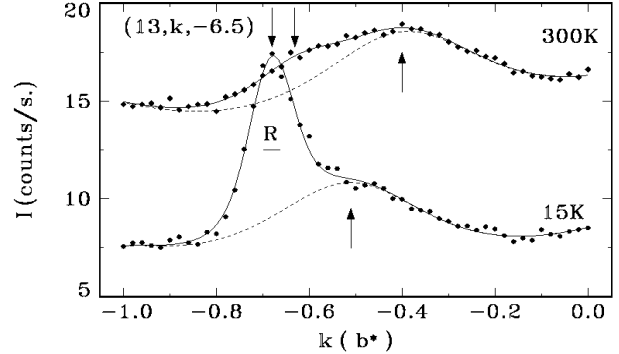,angle=90.0,height=5cm,width=8.5cm}}
\vskip 0.2cm
\caption{Profile of the $(13,-1+q_{b},-6.5)$ diffuse scattering at 300 K and
15 K. The downwards (upwards) arrows point towards the ${\bf q}_{c}$%
-scattering (${\bf q}_{s}$), respectively. The solid and dotted lines are
the results of fit described in the text. }
\end{figure}

This analysis is confirmed by 
the temperature variation of the satellite profile. Fig. 2. shows that the 
${\bf q}_{s}$ broad component is still present at 300 K, with the same 
intensity. The curves in fig. 2 are the result of a fit by a sum of 
Lorentzian-squared functions, convoluted to the resolution function. 
From this analysis, the temperature dependences of $1-q_{b}$ and $1-q_{s}$ 
have been extracted and are displayed in Fig. 3.

\begin{figure}
\centerline{\epsfig{file=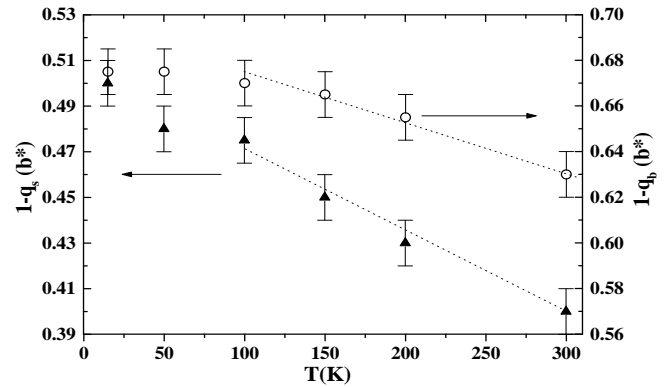,angle=90.0,height=5cm,width=8.5cm}}
\vskip 0.2cm
\caption{Temperature dependence of $1-q_{b}$ (right scale) and $1-q_{s}$
(left scale).}
\end{figure}

Another striking feature of Fig. 1 is the difference between the 
$+2{\bf k}_{F}$ and $-2{\bf k}_{F}$ profiles. Consistenly with the previous 
analysis, this has been interpreted as due to an intensity asymmetry 
$I_{a}^{\prime }$ of the ${\bf q}_{s}$-scattering, as indicated in Fig. 1.
At last, a comparison with the satellite reflection intensities of the pure
blue bronze (not shown here) reveals an intensity asymmetry of the 
${\bf q}_{c}$-scattering \cite{Rouz2}. 
This is evidenced by an intensity ratio ${I_{-2k_{F}}/I_{2k_{F}}}$ 20\% 
lower than  that of the pure compound at low temperature \cite{Rouz2}. 
This effet is indicated by $I_{a}$ in fig. 1.
 
As far as the low temperature CDW correlation 
lengths are concerned, they can be estimated from the inverse of 
the Half Widths at Half Maximum (HWHM) of the ${\bf q}_{c}$-scattering.
Along the chains, one finds $\ell _{b^{*}}=$28.7 \AA\ at 15 K \cite{L2}.
The ordered domains are quasi-one-dimensional and they contain on average 
one V substituant \cite{L2}. 

Let us now present a simple model, which indicates that the 
${\bf q}_{s}$-scattering is due to Friedel oscillations.
At $T=0$ K, the oscillating part of the electronic density of a metal 
at large distance $r$ from an impurity of charge -Z located at $r=0$ 
reads \cite{Ziman} : 
\begin{equation}
\delta \rho (r)\sim \frac{\cos (2k_{F}r+\eta )}{r^{D}},  \label{FOosc}
\end{equation}
where $D$ is the space dimension and $\eta$ is the phase shift of the
electronic wave functions at the Fermi level. With these definitions, 
$\eta$ is related to $Z$ through the Friedel sum rule  
$Z=\frac{2}{\pi }\eta$.

Physically, this relation means that the additional charge -Z has to be 
screened by the conduction electrons, by bringing an opposite charge Z in 
the vicinity of the impurity. In 1D, the lattice distortion associated to the 
FO reads : 
\begin{equation}
{\bf u}(x)= {\bf u}_{0}\frac{\exp (-\left| x\right| /\xi )}{\left| x\right| }%
\sin (2k_{F}x+\chi (x)).  \label{FOu}
\end{equation}
where $r=\left| x\right| $.
In this expression we have introduced a damping length $\xi $, due to the 
temperature and/or the disorder, and the phase function $\chi (x)$, whose
limits are $\chi (\pm \infty )=\pm \eta $, consistently with equation 
(\ref{FOosc}). $\chi (x)$ jumps between these values on a distance roughly 
equal to the extension of the potential of the impurity, {\it i.e.}
less than the lattice spacing.
At last, let us remark that a decrease of the distance between neighbors 
corresponds to a decrease (increase) of the hole (electron) density.
The two modulations are thus in quadrature, in agreement with expressions 
(\ref{FOosc}) and (\ref{FOu}).    

For the sake of simplicity, 
we shall model the x-ray scattering by taking the 1D lattice Fourier transform 
${\bf u}_{q}$ of ${\bf u}(x)$, which can be calculated exactly \cite{Rouz1}.
This 1D approximation is justified by the small tranverse correlation
lengths of the ${\bf q}_{s}$-scattering.
The intensity diffracted by the FO, given by the 
$I_{d}\sim \left| {\bf u}_{q}{\bf u}_{-q}\right|$ 
term, mainly depends on $\xi $ and $\eta $.  
Solid lines in fig. 4 represent the calculated intensity, for 
$\eta=\pm \pi /2$ and the blue bronze low temperature value of 
$2k_{F}=0.75b^{*}$.
Indeed, this value of $\eta$ corresponds to the screening
of a single charge impurity, which is the case here as discussed below.
For large values of $\xi $, the intensity exhibits a sharp discontinuity at 
$\pm 2k_{F}$ and a long tail in the small-q direction. These discontinuities 
are smoothed for smaller values of $\xi $, which slightly shifts the maxima 
from the $\pm 2k_{F}$ positions. As expected from the general analysis
\cite{Ravy2}, the phase distortion due to the FO phase shift gives rise to a 
profile asymmetry of the satellite reflections.

In fact, the total intensity diffracted by the FOs in the real crystal is given 
by (\ref{scatt}). As the Laue scattering is weak here, the only additional term 
is the third one : $I_{a}^{\prime }$. 
The dashed line of the lower part of Fig. 4 gives the total
scattering intensity in the $\Delta f \sin \eta <0$ case. 
As expected, an IA of the ${\bf +q/-q}$ is obtained, which gives the 
possibility of measuring $\sin \eta $. Indeed, using (\ref{scatt}), 
the IA term can be written: 
\begin{equation}
I_{a}^{\prime }({\bf Q}={\bf G}_{hkl}{\bf \pm q})\sim \pm \overline{f}\Delta
f({\bf Q.u}_{0})\sin \eta.  \label{sineta}
\end{equation}

\begin{figure}
\centerline{\epsfig{file=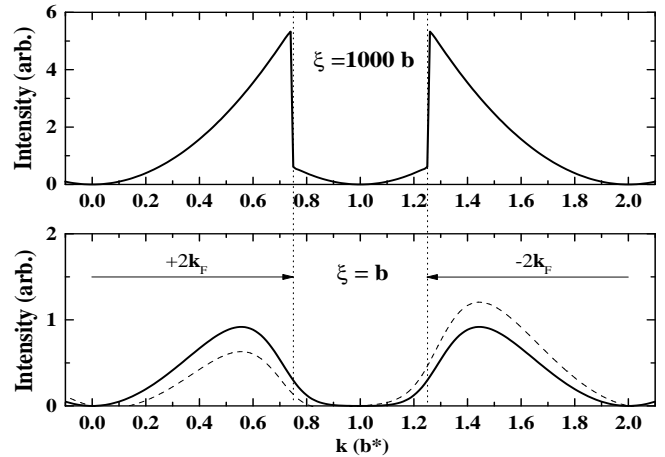,angle=90.0,height=6cm,width=8.5cm}}
\vskip 0.2cm
\caption{Intensity of the diffraction by a Friedel oscillation for $\xi $%
=1000 b (top) and $\xi $= b (bottom). The dashed curve is obtained by adding
the linear term $I_{a}^{\prime }$ for $\Delta f \sin \eta <0$. }
\end{figure}

A comparison of Figs. 1 and 4 strongly suggests the ${\bf q}_{s}$-scattering
to be due to FOs around the single charged V impurities. From the position
of the ${\bf q}_{s}$-scattering with respect to $2{\bf k}_{F}$, the damping
length can be estimated as $\xi \simeq 8$ \AA. The FOs are present at
ambient temperature, as shown in Fig. 2, and still exist in the CDW phase with 
the same characteristics. 
Concerning the ${\bf q}_{c}$-scattering, it corresponds to the
regular scattering by the $2k_{F}$ lattice distortion at larger distance from 
the impurity. 
The correlated variation of the ${\bf q}_{c}$- and the ${\bf q}_{s}$-scattering
maxima (Fig. 3), both related to the thermal variation of 
$2k_{F}$\cite{Girault}, supports this interpretation.

The observation of intensity asymmetries on both ${\bf q}_{c}$- and 
${\bf q}_{s}$-scattering gives evidence of a coherence between the position 
of an impurity, and respectively the FO and the CDW. 
Using eq. (\ref{cosphi}) and (\ref{sineta}) and the signs of $I_{a}$ 
and $I_{a}^{\prime }$ obtained from the experiment, one can readily calculate
the signs of $\cos \varphi _{0}$ and $\sin \eta$. Considering that 
$\Delta f=f_{V}-f_{Mo}<0$ and ${\bf Q.u}_{2{\bf k}_{F}}>0$ for positive 
k-values, one gets $\cos \varphi _{0} >0$ and $\sin \eta >0$, which
is consistent $\varphi _{0} \simeq 0$ and $\eta \simeq \frac{\pi }{2}$. 
Both values correspond to a local increase of
the hole density around the impurity, which is natural for the V substituant. 
Indeed, the V$^{5+}$ atom provide a negative charge (Z=1) with respect to the 
molybdenum Mo$^{6+}$ background. 
This charge has to be screened by a positive charge, which induces
a local increase of the hole density as found experimentally.
The $\eta \simeq \frac{\pi }{2}$ value is that expected from the Friedel sum 
rule with Z=1 

Fig. 5 displays a tentative representation of the CDW/FO structure around a
V atom according to the present experimental results. FO's dominate the
charge modulation in a tiny $2\xi \simeq$ 16 \AA\ region around the
impurity. Consistently with this result, this damping length can be
estimated at $\xi \simeq$ 9 \AA\ from $\xi^{-1} \simeq \ell^{-1} +
\xi_{0}^{-1}$, where $\xi _{0}=\hbar v_{F}/\Delta\simeq $ 13 \AA\ is the
electronic coherence length \cite{Schlenker}, and  $\ell=$ 27 \AA\ is the 
average distance beetween V substituants along the chains. 
At larger distances, the CDW dominates the charge oscillation. 
However, the CDW-phase extrapolated at the
impurity site is $\varphi_{0} \sim 0$, as indicated by the dotted line.
This description is consistent with the strong pinning picture as proposed 
in refs. \cite{Zawad}.
In addition, the region of mismatch between the CDW and the FO 
corresponds to the $q \simeq \xi^{-1}$ wave vectors, which indicates 
that the whole $\pm 2{\bf k}_F$ profiles contain also information on this 
important crossover region. 
Finally, let us point out that these $\pm 2{\bf k}_F$- asymmetric profiles, 
which clearly depends on FOs, CDWs and their interferences, are of great 
importance for the understanding of the sliding of CDWs and open a unique 
access to the properties of CDW at both microscopic (FO) and mesoscopic 
pinning) scales.

\begin{figure}
\centerline{\epsfig{file=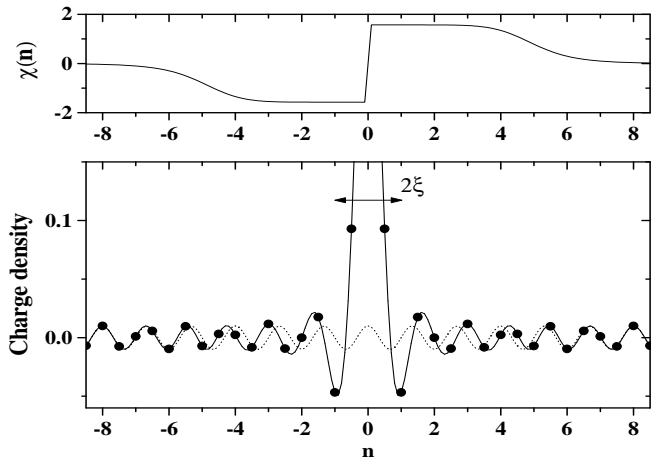,angle=90.0,height=6cm,width=8.5cm}}
\vskip 0.2cm
\caption{Bottom : sketch of the oscillating part of the hole charge density
around an impurity placed at the origin, as a function of the cell index $n$. 
Circles represent the Mo atoms. Top : Phase $\chi (x=nb)$ of
this charge density, in radian.}
\end{figure}

The conclusion of this study is the evidence of Friedel oscillations in the
vicinity of vanadium atoms in V-doped blue bronze. The observation of 
intensity asymmetries provides evidence of a coherence between the impurity
position and the CDW/FO. Moreover, the observation of the profile asymmetry
clearly indicates the presence of phase distortions, corresponding to a local
decrease of the electronic density. Additional experiments on crystal with
different doping levels are planned in order to elucidate more quantitatively
the microscopic features of the pinning of CDWs in low-dimensional materials.
More generally, we have demonstrated how methods of asymmetry 
analyses of the x-ray scattering allows one to study the subtle interplay
between structural and electronic properties in disordered systems. 

We are indebted to J.-F B\'{e}rar and R. Moret for their help during the
synchrotron radiation experiments.

\end{document}